\documentclass{article}
\usepackage{authblk}
\usepackage{setspace}
\usepackage{graphicx}
\usepackage{subcaption}
\usepackage{amsmath}
\usepackage{booktabs}
\usepackage{hyperref}
\usepackage{lineno}

\title{Immiscible two-phase flow in porous media: a statistical mechanics approach}
\author[1*]{Alex Hansen}
\author[1$\dag$]{Santanu Sinha}
\affil[1]{PoreLab, Department of Physics, Norwegian University of Science and Technology, N-7491 Trondheim, Norway}
\affil[*]{alex.hansen@ntnu.no}
\affil[$\dag$]{santanu.sinha@ntnu.no}
\date{\today}
\begin{document}
\maketitle
\begin{abstract}
The central problem in the physics of immiscible two-phase flow in porous media is to
find a proper description of the flow at scales large enough so that the medium may be
regarded as a continuum: the scale-up problem. So far, the only workable approach to
the multiphase flow scale-up problem has been a set of phenomenological equations that
have obvious weaknesses. Attempts at going beyond this relative permeability theory have
so far not led to practical applications due to exploding complexity. Edwin T.\ Jaynes
proposed in the fifties a generalization of statistical mechanics to non-thermal systems
based on the information theoretical entropy of Shannon. This approach is used to
construct a description of immiscible two-phase flow in porous media at the continuum
scales, which is directly related to the physics at the pore scale, and at a level of complexity that is manageable. The approach leads to a thermodynamics-like formalism at the
continuum scale with all the relations between variables that “normal” thermodynamics
has to offer. New emergent variables appear. Among these, the co-moving velocity
stands out as a key variable with implications for ordinary thermodynamics. We present 
here a short review of this approach.
\end{abstract}
\vspace{1em}
\section{Introduction}
\label{intro}

Porous media are as the name indicates, full of pores that are typically connected. More formally, one may define them as materials whose surface area is extensive in the volume.  This means that surface effects never go away no matter the scale at which they are studied.  Central to porous media research, both from a fundamental point of view, but also for applications, is the study of their transport properties.  One may study their fluid transport properties at the pore scale. One is then faced with a problem of hydrodynamics in confined spaces.  If there is more than one fluid present, all of them Newtonian, and they are immiscible, the problem becomes one of the viscous forces of the fluids competing with the capillary forces stemming from the fluid-fluid and fluid-matrix interfaces \cite{feder2022physics}.  

Considering the porous medium at a scale at which the pores are so small that it appears continuous, but not so large that it appears heterogeneous, fluid transport through it takes on a very different character. At this scale, known as the {\it Darcy scale,\/} concepts such as surface tension and interface dynamics no longer appear directly, but are worked into emergent properties that are observable on this scale. 

At the Darcy scale, the simultaneous flow of immiscible fluids can be described as the flow of a single non-Newtonian fluid whose local rheological properties are controlled by the transport of two scalar parameters, the saturations $S_w$ and $S_n$, which are linked by the relation
\begin{equation}
\label{eq1}
S_w+S_n=1\;.
\end{equation}
The saturations are the relative volume densities of the fluids with respect to the pore volume density. The subscripts $w$ and $n$ refer to the more wetting ($w$) of the two fluids and the less wetting ($n$) with respect to the solid matrix.  The viscosities of the two fluids $\mu_w$ and $\mu_n$ are parameters that survive moving from the pore scale to the Darcy scale. There is also a pressure $p$ and a fluid velocity $\vec{v}_p$, where the subscript ``p'' stands for ``pore''.       

Modeling immisicble two-phase flow in porous media at the Darcy scale has a long history in the physics literature.  The earliest paper we find is from 1934 by Muskat \cite{muskat1934two} where we find attempts at formulating constitutive equations for each of the two fluids.  Two years later, a complete formulation was in place \cite{wyckoff1936flow,muskat1936flow} --- an approach which today is known as relative permeability theory. It is a generalization of the fundamental work by Darcy in 1856 for single phase flow \cite{darcy1856fontaines}. A few years later (1941), Leverett introduced the  capillary pressure function, $p_c$ \cite{leverett1941capillary}.  This was necessary as the equations as they stood at that point, could not handle capillary suction (e.g.\ observed when dipping porous paper into water).  With his introduction of the capillary pressure curve, the theory took on its final form. It splits the Darcy scale fluid velocity $\vec v_p$ into a Darcy scale velocity for each of the two immiscible fluids, $\vec v_w$ and $\vec v_n$, 
\begin{equation}
\label{eq1-1}
\vec{v}_p=S_w\vec{v}_w+S_n\vec{v}_n\;.
\end{equation}
Hence, $\vec{v}_p$ is the saturation weighted average over the velocities of the two immiscible fluids. The theory also splits the pressure $p$ into pressures for each of the two fluids, $p_w$ and $p_n$. The difference between the two pressures is the Leverett capillary pressure $p_c$, so that
\begin{equation}
\label{eq2}
p_c=p_n-p_w\;.
\end{equation}
The generalized Darcy equations are
\begin{eqnarray}
\vec v_w &=& -\ \frac{Kk_{rw}}{\phi S_w\mu_w}\ \nabla p_w\;,\label{eq3}\\
\vec v_n &=& -\ \frac{Kk_{rn}}{\phi S_n\mu_n}\ \nabla p_n\;.\label{eq4}
\end{eqnarray}
Here $K$ is the permeability of the porous medium, a quantity measuring the conductivity of the porous medium with respect to the flow of a single fluid.  The porosity $\phi$ is the pore volume of an representative elementary volume divided by its volume. Both $K$ and $\phi$ are properties of the solid matrix and not the fluids. The two parameters $k_{rw}$ and $k_{rn}$ are the relative permeabilities. They reduce the effective permeability of the fluids due to the presence of the other fluid.

The central assumption in relative permeability theory is that the three parameters $p_c$, $k_{rw}$ and $k_{rn}$ only depend on the saturation $S_w$: $p_c=p_c(S_w)$, $k_{rw}=k_{rw}(S_w)$ and $k_{rn}=k_{rn}(S_w)$. 

Adding a continuity equation for the saturation,
\begin{equation}
\label{eq5}
\frac{\partial S_w}{\partial t}+\nabla\cdot (S_w \vec{v}_w) =0
\end{equation}
completes the description and we have a closed set of equations once the three parameters $p_c$, $k_{rw}$ and $k_{rn}$ have been determined.  Note here that the incompressibility of the two fluids is built into this equation. To see this, we note that we have a similar continuity equation for the non-wetting fluid,
\begin{equation}
\label{eq5-1}
\frac{\partial S_n}{\partial t}+\nabla\cdot (S_n \vec{v}_n) =0\;.
\end{equation}
We then add the two continuity equations and use the definition of the averaged fluid velocity $\vec{v}_p$ in Equation (\ref{eq1-1}) to find
\begin{equation}
\nabla\cdot\vec{v}_p=0\;,
\end{equation}
i.e., the incompressibility condition. 

Equations (\ref{eq2}) to (\ref{eq4}) are phenomenological, and the strong assumption that the three central parameters depend only on the saturation makes them easy to criticize. Nevertheless, they are completely dominating in all practical applications.

It is the aim of this short review to discuss a recent attempt at constructing a Darcy scale theory for immiscible two-phase flow in porous media that is built on upscaling, i.e.\ connecting the pore-scale physics to the Darcy scale. An in-depth review of the upscaling problem may be found in Berg et al.\ \cite{berg2026from}.  The theory we review here, based on the Jaynes generalization of statistical mechanics \cite{jaynes1957information,jaynes1957informationb}, have appeared in several papers, see \cite{hansen2018relations,hansen2023statistical,hansen2025thermodynamics} for some of the central ones.

Immiscible two-phase flow in porous media gives rise to a rich spectrum of behaviors.  We review these in the next section.  In particular, we present a Darcy-scale phase diagram in Figure \ref{fig3} showing that there is a glassy and a non-glassy phase separated by a critical line \cite{sinha2026glassy}. Section \ref{statmech} introduces the generalization of statistical mechanics to the immiscible two-phase flow problem in porous media, which then forms the background for the phase diagram of the previous section.  

Statistical mechanics in the context of the flow problem, scales up the pore-scale physics to the Darcy scale in the form of a thermodynamics-like formalism.  This is done in section \ref{darcy-thermodynamics}.  

We end this small review by some concluding remarks in section \ref{conclusion}. 

\begin{figure}[t]
    \centering
    \includegraphics[width=0.9\textwidth]{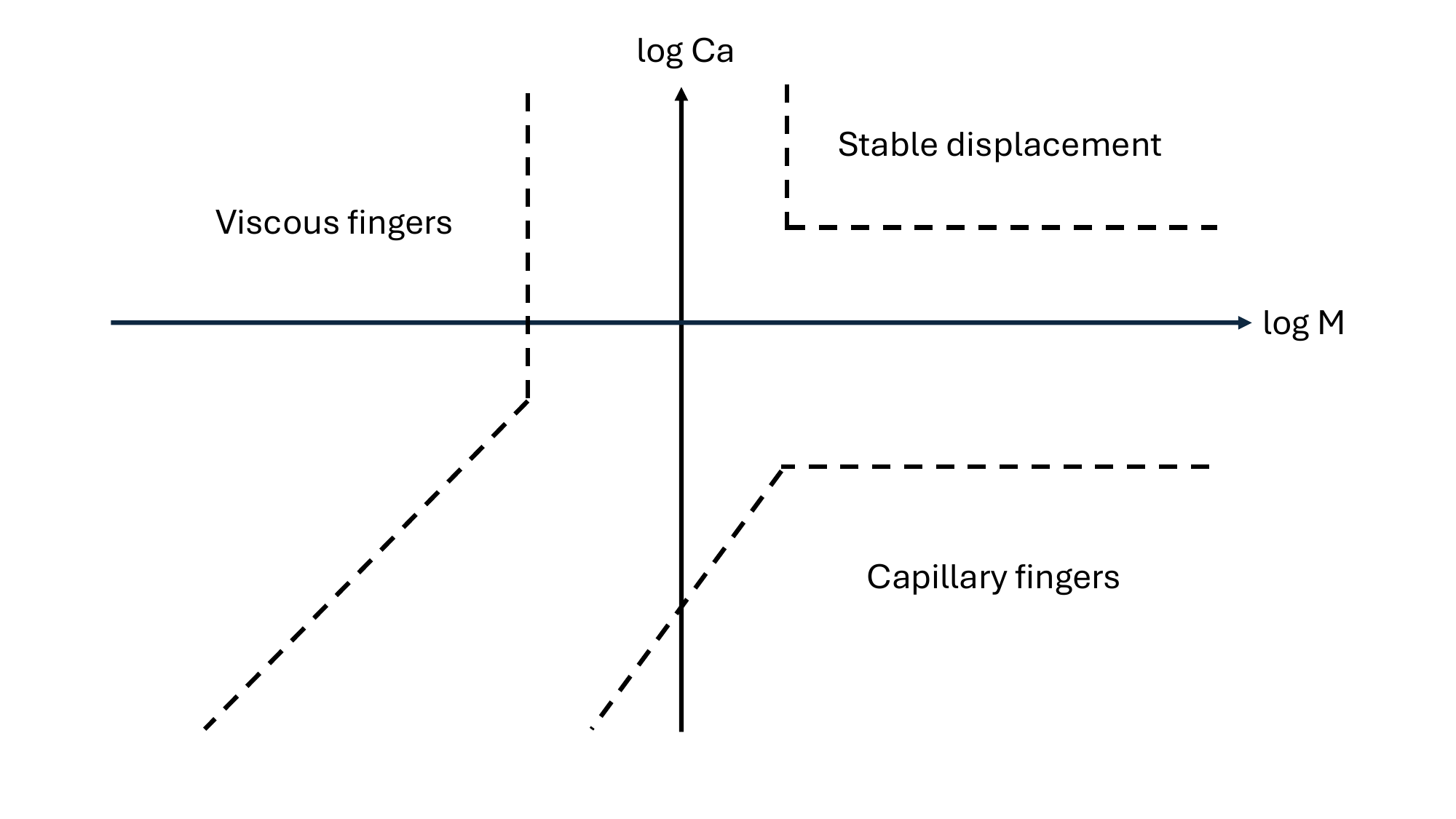}
    \caption{A sketch of the phase diagram that Lenormand et al.\ \cite{lenormand1988numerical} suggested for drainage processes, i.e., a less wetting fluid is injected into a porous medium saturated by a more wetting fluid with respect to the solid matrix. The axes are the logarithms of the viscosity ratio $M$ and the capillary number $Ca$, both defined in the text. There are three behaviors: viscous fingering, capillary fingering and stable displacement. There is a crossover region between them marked by the dashed lines.}
    \label{fig1}
\end{figure}

\section{Phase diagrams}
\label{phase_diagrams}

Already in 1988, Lenormand et al.\ introduced a two-dimensional pore-scale phase diagram for shapes that occur when a less wetting fluid is injected into a porous medium saturated by a more wetting fluid with respect to the solid matrix, i.e.\ drainage \cite{lenormand1988numerical}, see Figure \ref{fig1}. One axis is the logarithm of the non-dimensional viscosity ratio $M=\mu_w/\mu_n$.  The orthogonal axis is the logarithm of the non-dimensional capillary number $Ca=\mu_n v/\phi\sigma$.  Here $v$ is a typical velocity.  Under drainage, $v$ is the velocity of the injected fluid. Under steady-state flow conditions --- to be discussed in a moment --- we set $v=v_p$, the average velocity.  Furthermore, $\sigma$ is the interfacial tension between the two fluids.  Depending on the values of the two parameters $\log M$ and $\log Ca$, one would find viscous fingers, capillary finger or stable displacement patters occur.  Much of the work done within the physics community at that time in this field was focused on the geometry of these patterns.  This was the heyday of fractals \cite{feder1988fractals}. 

Drainage, and its counterpart imbibition (when a more wetting fluid is injected into a porous medium saturated with a less wetting fluid) are transients.  This is a much more complex situation to deal with compared to steady-state flow. Steady-state flow means continuously injecting simultaneously into the porous medium the two immiscible fluids.  One may think of a cylindrical porous sample, injecting the fluids at one end, and letting them leave at the other end by blocking off the side walls.  Inside this porous cylinder, away from the two edges, there will be {\it steady-state flow.\/} By this we mean that the Darcy scale variables will fluctuate around well-defined values. In particular, there will be no Darcy-scale gradients in the saturation.  It is this situation we will focus on in this review. 

\subsection{Pore scale phases}
\label{pore_scale_phases}

Avraam and Payatakes studied steady-state immiscible two-phase flow in model porous media \cite{avraam2006flow} at the pore scale.  They identified several types of flow patterns depending on the fluid parameters.  They defined the following pore-scale phases:
\begin{itemize}
\item{} {\it Continuous path flow\/} occurs when the flow is so slow that the interfaces between the fluids are held in place by capillary forces.  The flow is then along channels that are not obstructed by interfaces.  
\item{} {\it Large ganglion flow.\/}  A ganglion is a non-wetting fluid cluster which in this case is much larger than the pore size. These clusters move, break up and coalesce. There is considerable pressure difference internally in these ganglions, causing their motion.  This happens at higher capillary number than continuous path flow. 
\item{} {\it Small ganglion flow.\/} At even higher capillary number, the ganglions break down in size to being only somewhat larger than the pores. In this case, the pressure difference internally in the ganglions is small and they move due to the motion of the surrounding fluids. 
\item{} {\it Drop traffic flow.\/}  When the capillary number is raised even further, the ganglions break down into droplets inside the pores, leading to a situation similar to traffic flow in cities. Hence, the name.
\end{itemize} 

\subsection{Darcy scale phases}
\label{darcy_scale_phases}

We now turn to the Darcy-scale phase taxonomy of the same problem: immiscible steady-state two-phase flow in porous media. There are three parameters that define the flow at the Darcy scale, the capillary number $Ca$, the viscosity ratio $M$ and the saturation $S_w$.  Keeping $M$ and $S_w$ fixed, Berg et al.\ \cite{berg2026from} propose three phases that they name I, II and III in order of increasing $Ca$.  This was based on the constitutive equation relating flow velocity $v_p$ to the pressure gradient $|\nabla p|$.  

In 2009, Tallakstad et al.\ published two papers investigating experimentally steady-state immiscible two-phase flow in a model porous medium \cite{tallakstad2009steady,tallakstad2009steadyb}. At the capillary numbers this experiment was run at, they found a power law relation between the flow rate and pressure drop across the porous medium.  In terms of velocity vs.\ pressure gradient we would have
\begin{equation}
\label{eq6}
v_p=m_{\rm II} |\nabla p|^\beta\;,
\end{equation}
where $m_{\rm II}$ is the conductivity and $\beta=1.85\pm0.27$. This observation was followed up in a number of papers by different groups, see e.g.\ \cite{rassi2011nuclear,sinha2012effective,yiotis2013blob,aursjo2014film,sinha2017effective,gao2017x,yiotis2019nonlinear,roy2019effective,gao2020pore,zhang2020quantification,fyhn2021rheology,zhang2022nonlinear}. 

The following picture has emerged. At very low capillary numbers, the fluid interfaces held in place by the capillary forces essentially do not move or move little (I). This corresponds to the continuous pathway picture of Avraam and Payatakes \cite{avraam2006flow} and  gives rise to a linear relationship between velocity and pressure gradient,
\begin{equation}
\label{eq6-100}
v_p=m_{\rm I} |\nabla p|\;.
\end{equation}
As the capillary number is increases, the interfaces start moving, thus successively opening more and more flow channels. This causes the power law dependency in Equation (\ref{eq6}) (II).  At high enough capillary number, all the interfaces that potentially may move, are moving and as a consequence the system reverts to a linear relationship between velocity and pressure gradient (III), and we find again a linear relationship between fluid velocity and pressure gradient,
\begin{equation}
\label{eq6-200}
v_p=m_{\rm III} |\nabla p|\;.
\end{equation}

\begin{figure}[t]
    \centering
    \includegraphics[width=1.2\textwidth]{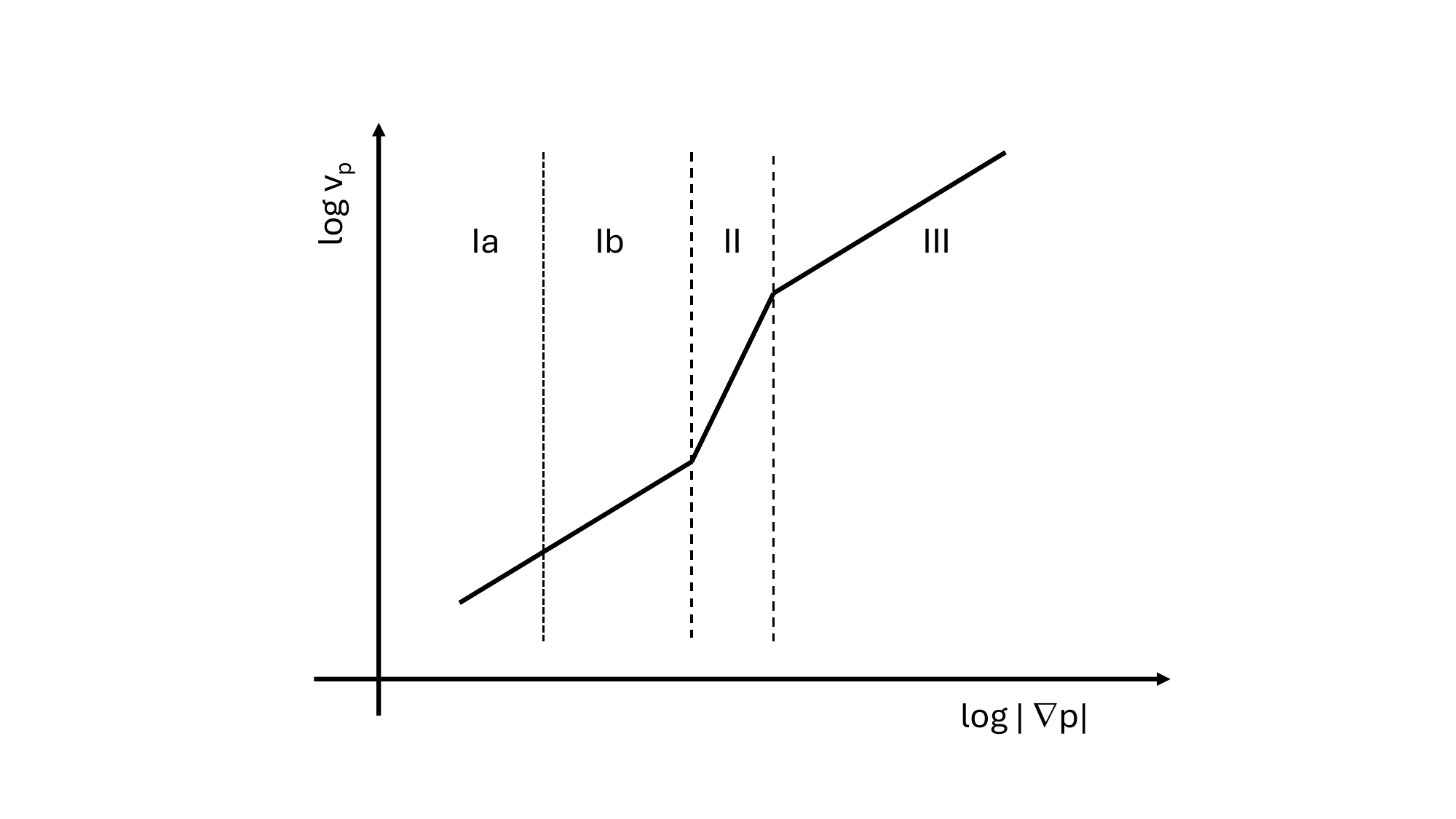}
    \caption{We illustrate the three constitutive Equations (\ref{eq6}) to (\ref{eq6-200}) and the phases Ia, Ib, II and III here. Phase Ia is characterized by the fluid interfaces being held in place by capillary forces and the fluid transport occurs through open channels.  In phase Ib, there is movement of the interfaces, but the flow is still dominated by flow through open channels. The relation between fluid velocity and pressure gradient is linear in phases Ia and Ib, see Equation (\ref{eq6-100}). Regime II is characterized by a power law relation between fluid velocity and pressure gradient (\ref{eq6}).  This is due to an increasing number of interfaces moving as the pressure gradient is increased.  Regime III is again linear, Equation (\ref{eq6-200}), due to all interfaces that can move are now moving.}
    \label{fig2}
\end{figure}

Based on the work of Armstrong et al.\ \cite{armstrong2016role}, Berg et al.\ \cite{berg2026from} split phase I into a phase Ia and a phase Ib.  They noted that before the power law behavior sets in, there is movement of the interfaces.  However, fluid transport is still dominated by the channels appearing in continuous path flow.  They named this phase Ib, and the phase where the interfaces are stuck, phase Ia. Gao et al.\  \cite{gao2020pore} described phase Ib in the following way: ``We observe the onset of dynamics, defined as when there is some intermittency but the Darcy law remains valid."
We summarize the phases Ia through III in Figure \ref{fig2}. 

\begin{figure}[t]
    \centering
    \includegraphics[width=0.7\textwidth]{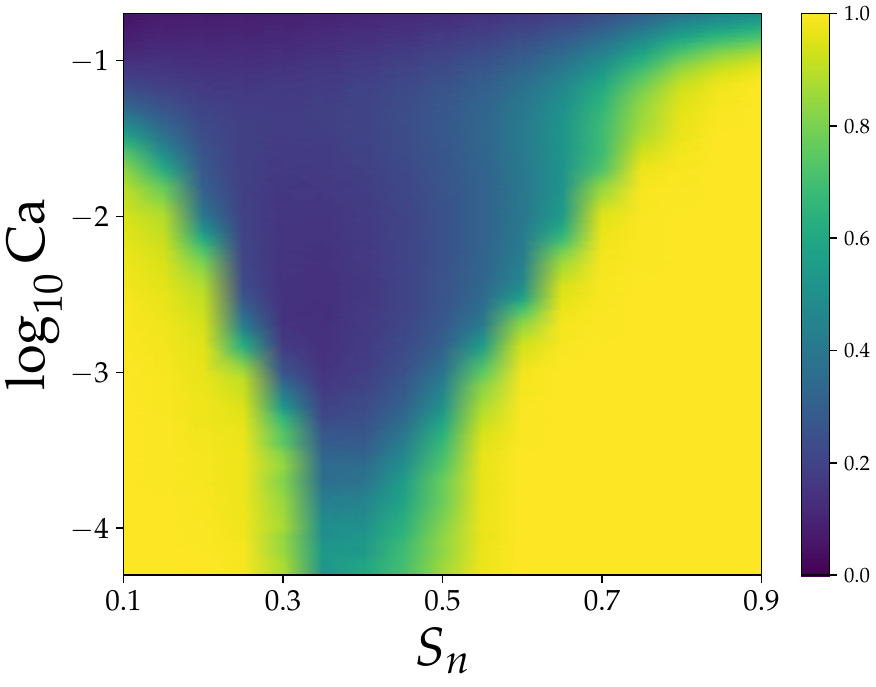}
    \caption{Phase diagram for immiscible two-phase flow in porous media for a viscosity ratio $M=1$.  The abscissa shows the non-wetting saturation $S_n$ and the ordinate $\log Ca$.  The color shows the value of the Edwards-Anderson order parameter, revealing a glassy phase and a non-glassy phase.  The transition line coincides with the onset of power law behavior, see Equation (\ref{eq6}). The figure is adapted from \cite{sinha2026glassy}.}
    \label{fig3}
\end{figure}

By mapping a dynamic pore network model \cite{sinha2021fluid} onto a spin model, and using Boltzmann machine learning \cite{meshulam2025statistical}, Sinha et al.\ \cite{sinha2026glassy} could identify a {\it glass transition\/}
where the Edwards-Anderson order parameter is zero at the high-capillary number side and larger than one on the low side, see Figure \ref{fig3}. Furthermore, the spin-glass susceptibility peaks \cite{binder1986spin} at the transition.  Remarkably, the transition coincides fully with the transition from linear to power law relation between velocity and pressure gradient Ib $\rightarrow$ II, see Figure \ref{fig2}. Phase Ib is therefore a {\it glassy flow state,\/} were there is motion of the interfaces. 

Going further down in capillary number past the glassy flow state, the system freezes, identifying phase Ia.  This is presumably a kinetic arrest \cite{foffi2000kinetic} and not a phase transition in the ordinary sense.  Its position will depend on the observation time of the system.  Likewise, the transition between the power law regime (\ref{eq6}) and the linear regime at high capillary numbers, i.e., where II $\rightarrow$ III in Figure \ref{fig2}, is also in the same way presumably not a phase transition in the ordinary sense. Roy et al.\ \cite{roy2024effective} have shown that the transition creeps towards the glassy transition when the system size increases.

Hence, only the glassy flow transition is a ``proper" phase transition --- or more precisely, a critical line as defined in thermodynamics. 

The glassy state then explains the hysteresis and the large fluctuations seen in phase Ib. The origin of the hysteretic behavior has been a theme for discussion in porous media research, with the presence of first-order transitions or missing variables being alternative explanations, see \cite{berg2026from}. 

\section{Non-thermal statistical mechanics}
\label{statmech}

We stated in the previous section that the glassy flow transition resembles a thermodynamic phase transition. The analysis by Sinha et al.\ \cite{sinha2026glassy} is performed under the assumption that the flow problem can be mapped onto an {\it equilibrium\/} system. But the flow problem is dissipative, and it is therefore not in equilibrium.  The way out of this apparent problem is to note that we are not scaling up molecular behavior to form ordinary thermodynamics. Rather, we are scaling up pore-scale hydrodynamics to the Darcy scale.  We are not dealing with ordinary molecular thermodynamics.  Rather, we will adapt the Jaynes generalized statistical mechanics approach \cite{jaynes1957information,jaynes1957informationb} to immiscible two-phase flow in porous media \cite{hansen2023statistical}. We follow the presentation in Hansen and Sinha \cite{hansen2025thermodynamics} fairly closely in the following.

We now sketch briefly the essence of the Jaynes generalized statistical mechanics. Central to his approach is to replace the Boltzmann statistical interpretation of thermodynamic entropy by the Shannon information entropy concept, being a quantitative measure of what is {\it not\/} known about a system. We consider a stochastic process producing $N$ possible outcomes, which we order from 1 to $N$, assuming a probability $p_i$ for the $i$th outcome, which is $x_i$.  A quantitative measure of our ignorance about this process must be independent of how we group events together. If we know nothing about the process, its information entropy must be at a maximum, since our ignorance is maximal.  This maximum is attained when all outcomes are equally probable, i.e., $p_i=1/N$.  There is only one function that can be constructed from $p_i$ with these properties, namely
\begin{equation}
\label{eq4-1}
S_I=-\sum_{i=1}^N p_i \ln p_i\;,
\end{equation}
which is then the information or Shannon entropy.  What happens if we {\it do\/} know something about the system, e.g., the average
\begin{equation}
\label{eq4-2}
\langle x\rangle =\frac{1}{N}\sum_{i=1}^N p_i x_i\;?
\end{equation}
Jaynes solution is to maximize the information entropy given the constraint (\ref{eq4-2}). This determines the probabilities $p_i$, which take the Gibbs form $p_i \propto \exp(-\lambda x_i)$. 

Jaynes provided a set of circumstances for his generalized statistical mechanics to be applicable. Hansen et al.\ \cite{hansen2023statistical} demonstrates that these circumstances are met for steady-state immiscible two-phase flow in porous media, and hence proceeded to construct such a statistical mechanics. Even if {\it molecular\/} entropy is produced, the pore-scale Shannon entropy is {\it not\/} produced under steady-state flow.  Hence, we may treat the system as being in equilibrium.   
\begin{figure}
    \centering
    \includegraphics[width=0.5\textwidth]{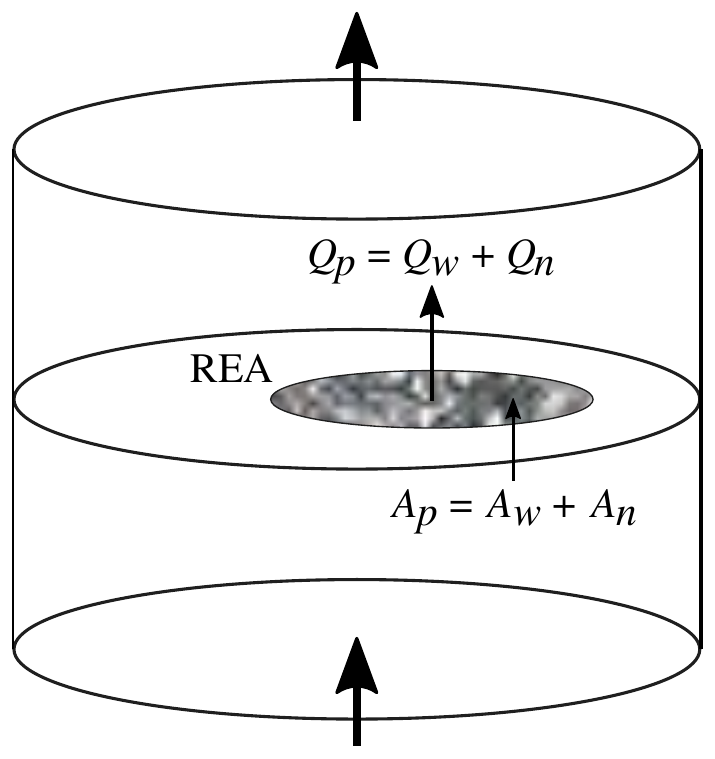}
    \caption{We define a Representative Elementary Area (REA) within a cut orthogonal to the average flow direction through the cylinder-shaped porous medium sample.  There is a wetting fluid flow rate $Q_w$ and a non-wetting fluid flow rate $Q_n$ passing through the REA. The total flow rate is $Q_p$. The wetting fluid covers an area $A_w$ of the REA and the non-wetting fluid an area $A_n$. The total pore area of the REA is $A_p$. Adapted from \cite{hansen2025thermodynamics}.\label{fig4}}
\end{figure}

We consider a cylindrical pore sample as shown in Figure \ref{fig4}.  Two immiscible fluids enter at the bottom and leave at the top.  The side walls are closed off. We assume that the pore structure and chemical composition of the matrix to be statistically uniform throughout the cylinder so that away from the edges, steady-state flow is found. If we make transversal cuts through the sample, the structure of the porous medium and the distribution of fluids in the pore space to be statistically uniform in this region: Comparing one cut to another, we will not be able to determine which is closest to the inlet.  Figure \ref{fig4} shows one such cut.  Within this cut, we pick out a {\it Representative Elementary Area\/} (REA), which is large enough to represent the statistics of the sample and small enough to have the flow fluctuate.  There is a volumetric flow rate $Q_p$ passing through the REA which may be split into a volumetric flow rate of the wetting fluid $Q_w$ and the volumetric flow rate of the non-wetting fluid $Q_n$, i.e.,
\begin{equation}
\label{eq2-1}
Q_p=Q_w+Q_n\;.
\end{equation}

The area of the REA is $A$. An area $A_p<A$ cuts through pore space.\footnote{A note on the notation: The ``p" in $Q_p$ refers to the pressure gradient being a control variable, whereas in $A_p$, the ``p" refers to ``pore."We keep this notation in order to facilitate access to the older literature.}  This area may be divided into areas cutting through the pore space filled with the more wetting fluid, $A_w$ and though the pore space filled with the less wetting fluid, $A_n$. We have that 
\begin{equation}
\label{eq3-1}
A_p=A_w+A_n\;.
\end{equation}

At a given time for a given REA, the fluid configuration, characterized by the velocity field $\vec v_p$ and where the fluids are, is $X$. The quantities that describe the REA depend on $X$, so that we have $Q_p(X)$, $Q_w(X)$, $Q_n(X)$, $A_p(X)$, $A_w(X)$, and $A_n(X)$, with $Q_n(X) = Q_p(X)-Q_w(X)$ and $A_n(X) = A_p(X)-A_w(X)$.  $A_p(X)$ does not dependent on the velocity field, nor the distribution of the two fluids, but only on the shape of pore space.  The spatial statistical fluctuations of $A_p$ are quenched, whereas the temporal and spatial fluctuations of the fluids are annealed.  We denote $p(X)$ the probability density to find fluid and pore space configuration $X$. We define a {\it configurational entropy,\/}
\begin{equation}
\label{eq4-3}
S=-\int dX\ p(X)\ \ln p(X)\;.
\end{equation}
We note that this is a differential entropy as the probability distribution $p(X)$ is continuous \cite{floerchinger2020thermodynamics}.

We assume that we know the averages (in time and position of the REA) of the variables just described, i.e.,
\begin{eqnarray}
Q_p&=&\int dX\ p(X)\ Q_p(X)\;,\label{eq4-4}\\
A_w&=&\int dX\ p(X)\ A_w(X)\;,\label{eq4-5}\\
A_p&=&\int dX\ p(X)\ A_p(X)\;,\label{eq4-6}
\end{eqnarray}
are known.

Following the Jaynes maximum entropy principle, we maximize the entropy (\ref{eq4-3}) given the constraints (\ref{eq4-4}) to (\ref{eq4-6}), finding that
\begin{equation}
\label{eq4-7}
p(X;\lambda_u,\lambda_w,\lambda_p)=\frac{1}{Z(\lambda_u,\lambda_w,\lambda_p)}\ \exp\left[-\lambda_u Q_p(X)-\lambda_w A_w(X)-\lambda_p A_p(X)\right]\;,
\end{equation}
where 
\begin{equation}
\label{eq4-8}
Z(\lambda_u,\lambda_w,\lambda_p)=\int\ dX\ \exp\left[-\lambda_u Q_p(X)-\lambda_w A_w(X)-\lambda_p A_p(X)\right]
\end{equation} 
is the partition function. Three new variables have appeared, $\lambda_u$, $\lambda_w$, and $\lambda_p$. 
We rewrite the partition function as
\begin{equation}
\label{eq4-13}
Z(\lambda_u,\lambda_w,\lambda_p)=\exp\left[-\lambda_u Q_z(\lambda_u,\lambda_w,\lambda_p)\right]\;.
\end{equation}

Ordinary statistical mechanics is an upscaling procedure for taking a molecular description of a system to the continuum scale, i.e., thermodynamics.  The generalized statistical mechanics we have just sketched here does the same thing: It takes the flow problem from the pore scale where $Q_p(X)$, $A_p(X)$ and $A_w(X)$ are defined to the Darcy scale, where a formalism similar to thermodynamics occurs.  

\section{Non-thermal thermodynamics at the Darcy scale}
\label{darcy-thermodynamics}

We now explore this non-thermal ``thermodynamics" appearing as the end-result of the generalized statistical mechanics described in the previous section.

It is convenient to define new variables based upon the Langrange mulitpliers $\lambda_u$, $\lambda_w$ and $\lambda_p$, 
\begin{eqnarray}
\theta&=&+\frac{1}{\lambda_u}\;,\label{eq4-15}\\
\mu&=&-\frac{\lambda_w}{\lambda_u}\;,\label{eq4-161}\\
\pi&=&-\frac{\lambda_p}{\lambda_u}\;.\label{eq4-171}
\end{eqnarray}

By using the definitions of $S$, $A_w$ and $A_p$, Equations (\ref{eq4-3}), (\ref{eq4-5}) and (\ref{eq4-6}), combined with the partition function $Z$ in Equation (\ref{eq4-8}) and the flow rate $Q_z$ defined in Equation (\ref{eq4-13}), we find
\begin{eqnarray}
S&=&-\left(\frac{\partial Q_z}{\partial \theta}\right)_{\mu,\pi}\;,\label{eq4-9}\\
A_w&=&-\left(\frac{\partial Q_z}{\partial \mu}\right)_{\theta,\pi}\;,\label{eq4-16}\\
A_p&=&-\left(\frac{\partial Q_z}{\partial \pi}\right)_{\theta,\mu}\;.\label{eq4-17}
\end{eqnarray}
The flow rate $Q_p$ is given by
\begin{equation}
\label{eq4-31}
Q_p(\theta,A_w,A_p)=Q_z(\theta,\mu,\pi)+\mu A_w+\pi A_p\;.
\end{equation}
This equation may be recognized as a Legendre transform (assuming convexity of the involved functions) from the variables $(\theta,\mu,\pi)$ to the variables $(S,A_w,A_p)$.

Let us also define a flow rate $Q_u=Q_u(\theta,A_w,A_p)$ given by 
\begin{equation}
\label{eq4-560}
Q_u(S,A_w,A_p)=Q_p(\theta,A_w,A_p)-S\theta\;.
\end{equation}
We also define a flow rate $Q_g(\theta,\mu,A_p)$ as
\begin{equation}
\label{eq4-562}
Q_g(\theta,\mu,A_p)=Q_z(\theta,\mu,\pi)+\pi A_p\;.
\end{equation}

We see that $\theta$ (Equation (\ref{eq4-15})) plays the same role as temperature in ordinary thermodynamics.  Hansen et al.\ \cite{hansen2023statistical} named it {\it agiture,\/} which is short hand for ``agitation temperature."  The variable $\mu$ (Equation (\ref{eq4-161})) is somewhat reminiscent of chemical potential, was named the {\it flow derivative.\/} The variable $\pi$, defined in Equation (\ref{eq4-171}), they called the {\it flow pressure.\/} It has no equivalent in ordinary thermodynamics. 

How should we understand the difference between the volumetric flow rates $Q_u$ and $Q_p$?  The flow rate $Q_p$ is that we inject into our porous sample at one end and which comes out on the other end.  It is a conserved quantity throughout the sample as any cut through the sample will have the same $Q_p$ passing through it. But locally there are fluctuations in the flow rate due to the competition between the capillary and viscous forces.  These fluctuations average to zero by construction for $Q_p$. $Q_p$ is based on sampling the flow configurations when the control parameters $(\theta,A_w,A_p)$ are fixed.  When our sampling assumes the control parameters $(S,A_w,A_p)$ fixed, these fluctuations no longer average to zero as we are averaging in a different way, making $Q_u$ different from $Q_p$.       

This is completely analogous to the difference between the internal energy (corresponding to $Q_u$) and the Helmholtz free energy (corresponding to $Q_p$) in ordinary thermodynamics.  The internal energy contains a fluctuating part due to thermal agitation that is not recoverable in terms of work, whereas the free energy is the energy free to do work --- i.e.\ the part in the internal energy due to fluctuations are subtracted off.   In the flow problem, $Q_u$ contains both the recoverable flow rate, i.e.\ the flow rate coming out of the sample and a part due to the fluctuations.  $Q_p$ is the free flow rate, i.e. the recoverable flow rate. 

Returning to the REA shown in Figure \ref{fig4}, we see that there will be fluctuations in the entropy $S$ and the wetting area $A_w$ --- but not in the pore area $A_p$.  The natural ``free" flow rate in this case is then $Q_g(\theta,\mu,A_p)$, which includes both of these types of fluctuations.  If we now in addition average over the position of the REA, we need to include the fluctuations in $A_p$, making $Q_z(\theta,\mu,\pi)$ the natural flow rate.   

The variables $Q_u$, $Q_z$, $Q_p$, $A_w$, $A_p$, and $S$ are all extensive in the area of the REA. The emergent variables $\theta$, $\mu$, and $\pi$ are intensive. We rescale $Q_p$, $A_w$, and $A_p$ by $A$, defining
\begin{eqnarray}
\phi&=&\frac{A_p}{A}\;,\label{eq5-1}\\
S_w&=&\frac{A_w}{A_p}=\frac{A_w}{A\phi}\;,\label{eq5-2}\\
v_p&=&\frac{Q_p}{A_p}=\frac{Q_p}{A\phi}\;.\label{eq5-3}
\end{eqnarray}

In order to keep track of what happens to the variables in, e.g., $Q_p(\theta,A_w,A_p,A)$ (where we now write the area $A$ explicitly), we express the extensivity through the scaling relation 
\begin{equation}
\label{eq5-4}
\lambda Q_p(\theta,A_w,A_p,A)=Q_p(\theta,\lambda A_w,\lambda A_p,\lambda A)\;.
\end{equation}
 We set the scale factor $\lambda=1/A_p$, finding the average fluid velocity to be
\begin{equation}
\label{eq5-5}
v_p\left(\theta,S_w,\phi\right)=\frac{1}{A_p}Q_p\left(\theta,A_w,A_p,A\right)=Q_p\left(\theta,S_w,1,\frac{1}{\phi}\right)\;.
\end{equation}

The three emergent variables $\theta$, $\mu$, and $\pi$,  are the conjugate variables to the flow rate $Q_u$, the wetting area $A_w$, and the pore area $A_p$, respectively.  We now go on to discuss further the meaning of these variables.  

\subsection{Agiture}
\label{agiture}

\begin{figure}[t]
    \centering
    \includegraphics[width=0.3\textwidth]{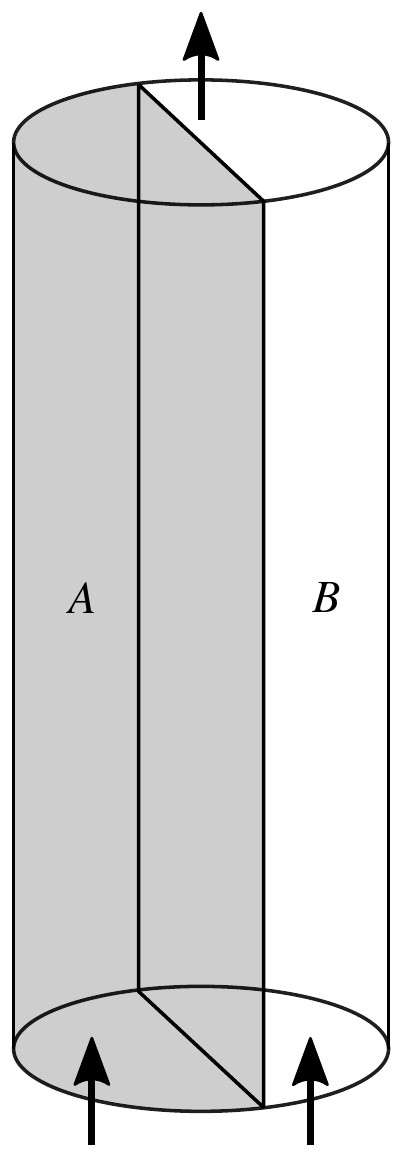}
    \caption{A porous cylinder where we inject the two immiscible fluids at the bottom.  The cylinder consists of two halves named A and B.  The difference between the two halves may e.g.\ be in chemical composition of the matrix, or they may consist of differently structured pore spaces.  However, we assume each of the two halves each to be statistically homogeneous.  Adapted from \cite{hansen2025thermodynamics}.}
    \label{fig5}
\end{figure}

We consider a porous cylinder as in Figure \ref{fig4}, but with the difference that it is divided into two halves A and B having different properties as shown in Figure \ref{fig5}.  The difference between the two halves may e.g.\ be in chemical composition of the matrix, or it may consist of differently structured pore spaces.  However, we assume the two halves each to be statistically homogeneous. We now assume that two immiscible and incompressible fluids are simultaneously injected into the cylinder under steady-state conditions.  

The flow configurational entropy $S$ remains constant under such conditions. As shown by Hansen et al.\ \cite{hansen2023statistical}, this means that the agiture in sectors A and B, $\theta^A$ and $\theta^B$, are equal,
\begin{equation}
\label{eq6-1}
\theta^A=\theta^B\;.
\end{equation}

But, what {\it is\/} the agiture?  In the example just given, each of the two zones will each have a pressure gradient $\nabla p^A$ and a pressure gradient $\nabla p^B$ driving the flow.  The average flow direction in both halves will be along the cylinder axis.  Hence, we have that 
\begin{equation}
\label{eq6-19}
\nabla p^A=\nabla p^B\;.
\end{equation}
It is natural to assume the relation
\begin{equation}
\label{eq6-21}
\theta=c |\nabla p|\;,
\end{equation}
when comparing Equations (\ref{eq6-1}) and (\ref{eq6-19}). Intuitively, this makes sense as the higher the pressure gradient, the higher the agiture.

We note that $Q_p=Q_p(\theta,A_w,A_p,A)$, defined in Equation (\ref{eq4-31}), may be written as
\begin{equation}
\label{eq6-6}
Q_p=Q_p(p',A_w,A_p,A)=A_p v_p(p',S_w,\phi)\;,
\end{equation}
where we have also used Equation (\ref{eq5-5}). We have set $p'=|\nabla p|$. Hence, this is a constitutive equation relating average fluid velocity to the local pressure gradient, saturation, and porosity.              

Hansen et al.\ \cite{hansen2025thermodynamics} argue that also the flow derivative, Equation (\ref{eq4-161}), is uniform under steady-state flow,
\begin{equation}
\label{eq6-2}
\mu^A=\mu^B\;.
\end{equation}

No such relation exists, however, for the flow pressure $\pi$, defined in Equation (\ref{eq4-171}). As already pointed out, the disorder associated with $A_p$, the conjugate of $\pi$, is quenched.  The proper way to deal with this disorder is to follow the recipe used for glasses \cite{binder1986spin}. Sinha et al.\ \cite{sinha2026glassy} is a step in this direction. 

\subsection{Thermodynamic velocities and the co-moving velocity}
\label{v-hat}

Hansen et al.\ \cite{hansen2018relations} defined two {\it thermodynamic velocities\/}
\begin{eqnarray}
{\hat v}_w=\left(\frac{\partial Q_p}{\partial A_w}\right)_{\theta,A_n}\;,\label{eq6-14}\\
{\hat v}_n=\left(\frac{\partial Q_p}{\partial A_n}\right)_{\theta,A_w}\;,\label{eq6-15}
\end{eqnarray}
where the control variables are $\theta$, $A_w$, and $A_n$, defined in (\ref{eq3-1}), making $A_p$ a dependent variable. Changing the variables $(\theta,A_w,A_n)\to(\theta,S_w,A_p)$, these two equations may be written as
\begin{eqnarray}
{\hat v}_w=v_p+S_n\left(\frac{\partial v_p}{\partial S_w}\right)_{\theta,\phi}\;,\label{eq6-16}\\
{\hat v}_n=v_p-S_w\left(\frac{\partial v_p}{\partial S_w}\right)_{\theta,\phi}\;.\label{eq6-17}
\end{eqnarray}

The flow derivative $\mu$, which is the conjugate of the wetting area $A_w$, and thereby also the saturation $S_w$, is given by
\begin{equation}
\label{eq6-18}
\mu=\left(\frac{\partial Q_p}{\partial A_w}\right)_{\theta,A_p}=\left(\frac{\partial v_p}{\partial S_w}\right)_{\theta,\phi}\;.
\end{equation}
Hence, Equation (\ref{eq6-17}) may be recognized as a Legendre transformation substituting $S_w \to \mu$, 
\begin{equation}
\label{eq6-191}
{\hat v}_n(\theta,\mu,\phi)=v_p(\theta,S_w,\phi)-S_w(\theta,\mu,\phi)\mu\;.
\end{equation}
In other words, the non-wetting thermodynamic velocity is the Legendre transformation of the average velocity with respect to the saturation. 

If we combine Equations (\ref{eq4-31}) and (\ref{eq4-562}), we find
\begin{equation}
\label{eq7-562}
Q_g(\theta,\mu,A_p)=Q_p(\theta,A_w,A_p)-A_w\mu\;.
\end{equation}
By comparing this expression to Equation (\ref{eq6-191}), we see that there is a direct association between the ``free'' flow rate $Q_g$ to the non-wetting thermodynamic velocity $\hat{v}_n$, namely 
\begin{equation}
\label{eq7-563}
Q_g=A_p \hat{v}_n\;.
\end{equation}

The thermodynamic velocities defined in Equations (\ref{eq6-14}) and (\ref{eq6-15}) are not the average velocities for each of the two fluid species, which we define as 
\begin{eqnarray}
{v}_w=\frac{Q_w}{A_w}\;,\label{eq7-1}\\
{v}_n=\frac{Q_n}{A_n}\;.\label{eq7-2}
\end{eqnarray}
They are the velocities that appeared in generalized Darcy Equations (\ref{eq2}) and (\ref{eq3}), and
they are the velocities that are measured in the laboratory, whereas the thermodynamic velocities are not. 

We rewrite Equation (\ref{eq2-1}) as
\begin{equation}
\label{eq7-10}
Q_p=Q_w+Q_n=v_wA_w+v_nA_n\;.
\end{equation}
The saturations in terms of the variables $A_w$ and $A_n$ are given by 
\begin{eqnarray}
S_w&=&\frac{A_w}{A_w+A_n}\;,\label{eq7-12}\\
S_n&=&\frac{A_n}{A_w+A_n}\;,\label{eq7-13}
\end{eqnarray}
leading to Equation (\ref{eq1}). We may then write Equation (\ref{eq7-10}) as
\begin{equation}
\label{eq7-15}
v_p=v_wS_w+v_nS_n\;,
\end{equation}
where $v_p$, the average velocity, is defined in Equation (\ref{eq5-5}).

We now turn to the thermodynamic velocities, Equations (\ref{eq6-14}) and (\ref{eq6-15}). The average flow rate $Q_p$ obeys the 
Euler scaling relation,
\begin{equation}
\label{eq7-16}
\lambda Q_p(p',A_w,A_n)=Q_p(p',\lambda A_w,\lambda A_n)\;.
\end{equation}
We then have from the Euler theorem for homogeneous functions,
\begin{equation}
\label{eq7-17}
Q_p=\left(\frac{\partial Q_p}{\partial A_w}\right)_{A_n,\phi,p'}A_w+\left(\frac{\partial Q_p}{\partial A_n}\right)_{A_w,\phi,p'}A_n=\hat{v}_wA_w+\hat{v}_nA_n\;.
\end{equation}
Dividing this equation by $A_p$ gives
\begin{equation}
\label{eq7-18}
v_p={\hat v}_w S_w+{\hat v}_n S_n\;.
\end{equation}
Comparing Equations (\ref{eq7-15}) and (\ref{eq7-18}) we find
\begin{equation}
\label{eq7-19}
v_p=v_wS_w+v_nS_n={\hat v}_w S_w+{\hat v}_n S_n\;.
\end{equation}
This equation does not imply that $v_w={\hat v}_w$ and $v_n={\hat v}_n$.  Rather, the most general relation between the thermodynamic and seeping velocities are 
\begin{eqnarray}
v_w={\hat v}_w-S_n v_m\;,\label{eq7-20}\\
v_n={\hat v}_n+S_w v_m\;,\label{eq7-21}
\end{eqnarray}
where $v_m$ is the {\it co-moving velocity\/} \cite{berg2026from,hansen2018relations,roy2020flow,roy2022co,pedersen2023parametrizations,alzubaidi2024impact,hansen2024linearity}.  We may combine these two equations with Equations (\ref{eq6-16}) and (\ref{eq6-17}) to find 
\begin{eqnarray}
v_w&=&v_p+S_n\left[\left(\frac{\partial v_p}{\partial S_w}\right)_{\phi,p'}-v_m\right]\;,\label{eq7-22}\\
v_n&=&v_p-S_w\left[\left(\frac{\partial v_p}{\partial S_w}\right)_{\phi,p'}-v_m\right]\;.\label{eq7-23}
\end{eqnarray}
By taking the derivative of Equation (\ref{eq7-19}) with respect to $S_w$, and using {Equations (\ref{eq7-20}) and (\ref{eq7-21})}, we find 
\begin{equation}
\label{eq2-23} 
v_m=\left(\frac{\partial v_p}{\partial S_w}\right)_{\phi,p'}-v_w+v_n=S_w\left(\frac{\partial v_w}{\partial S_w}\right)_{\phi,p'}+S_n\left(\frac{\partial v_n}{\partial S_w}\right)_{\phi,p'}\;.
\end{equation}

\begin{figure}
    \centering
    \includegraphics[width=1.\textwidth]{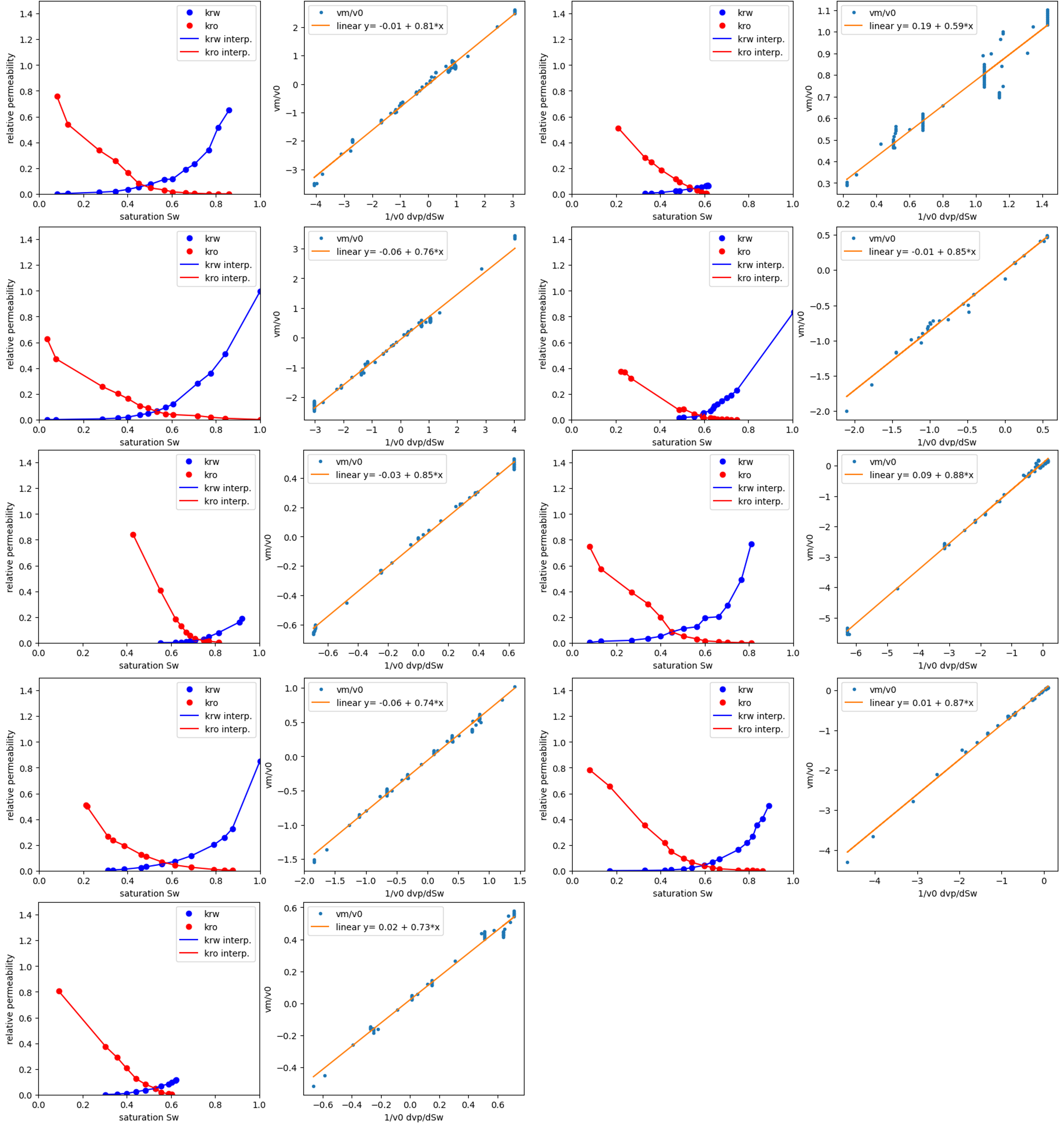}
    \caption{Measured relative permeability curves, $k_{rw}(S_w)$ and $k_{rn}(S_w)$ plotted against $S_w$, and the corresponding co-moving velocity $v_M/v_0$ plotted against $(1/v_0)dv_p/dS_w= \mu/v_0$, where $v_0$ is a velocity scale. From \cite{berg2026from}.\label{fig6}}
\end{figure}

We now return to Equation (\ref{eq6-191}), which may now be expressed in terms of the velocity $v_n$, rather than the thermodynamic velocity ${\hat v}_n$, 
\begin{equation}
\label{eq7-24}
v_n(\theta,\mu,\phi)-S_w(\theta,\mu,\phi)v_m(\theta,\mu,\phi)=v_p(\theta,S_w,\phi)-S_w(\theta,\mu,\phi)\mu\;,
\end{equation}
signifying that $(\theta,\mu,\phi)$ are the natural variables for the co-moving velocity. Measurements of the co-moving velocity suggest that it has the simple functional form \cite{berg2026from,roy2022co,alzubaidi2024impact,hansen2024linearity},
\begin{equation}
\label{eq7-25}
v_m(\theta,\mu,\phi)=a(\theta,\phi)+b(\theta,\phi)\mu\;,
\end{equation}
based on analysis of relative permeability data, on dynamic pore network modeling, and on lattice Boltzmann simulations on reconstructed sandstones We show Figure \ref{fig6} the co-moving velocity vs. $\mu$ based on a large number of relative permeability data sets.  

The thermodynamic velocities $({\hat v}_w,{\hat v}_n)$ can be found by knowing $v_p$ alone; see Equations (\ref{eq6-16}) and (\ref{eq6-17}).  The velocities, on the other hand, requires additional knowledge, i.e., $v_p$ and $v_m$, which are given by Equations (\ref{eq7-22}) and (\ref{eq7-23}). They provide a two-way mapping,  
\begin{equation}
\label{eq7-26}
\begin{pmatrix}
v_w\\
v_n\\
\end{pmatrix}
\leftrightarrow
\begin{pmatrix}
v_p\\
v_m\\
\end{pmatrix}
\;.
\end{equation}

\subsection{Is there an equivalent to the co-moving velocity in ordinary thermodynamics?}
\label{co-molar}

The answer to the question posed in the title of this section is ``yes." In fact, for any extensive variable, i.e., {\it additive\/} variable for two-component mixtures, there will be a corresponding co-``something" function.\footnote{Note that the capillary pressure (\ref{eq2}) and the generalized Darcy equations (\ref{eq3}) and (\ref{eq4}) split an {\it intensive\/} variable, the pressure, into two, $p\to (p_w,p_n)$. This is problematic, and the capillary pressure function is indeed problematic, both to measure and to handle.} 

\begin{figure}
    \centering
    \includegraphics[width=1.\textwidth]{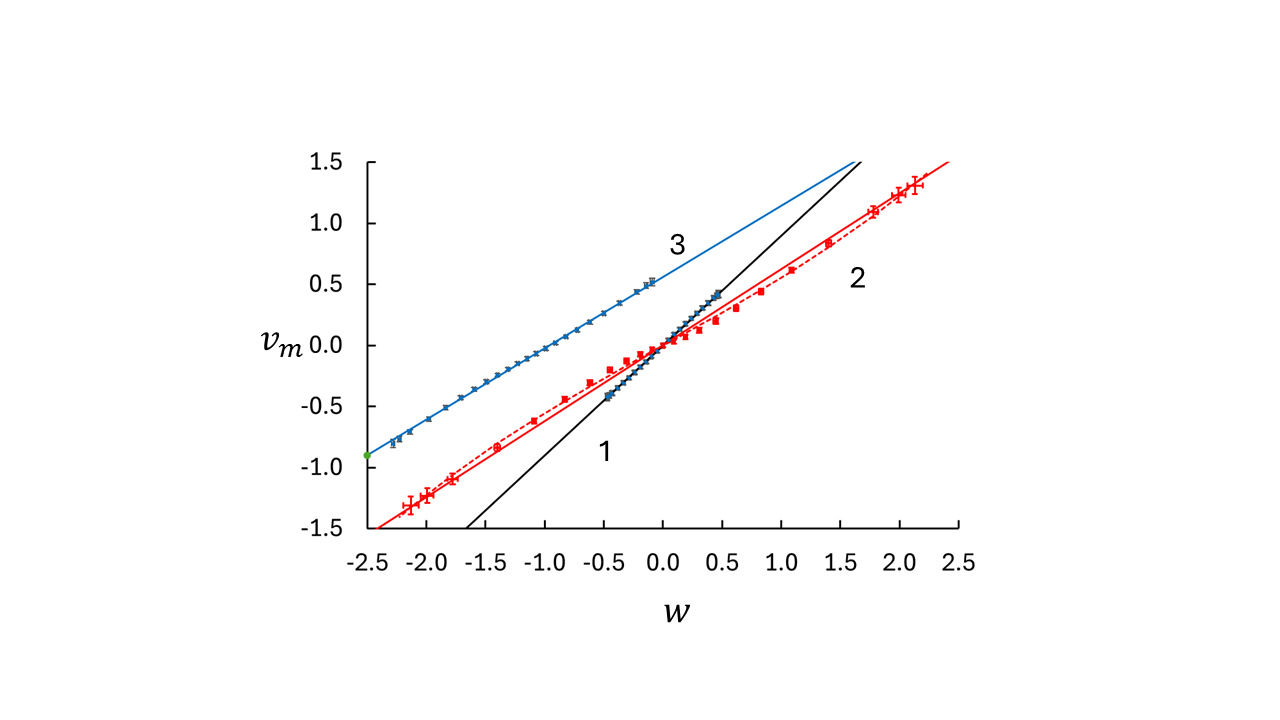}
    \caption{Co-molar volume ($\nu_m$) as a function of partial molar volume difference ($w = d\nu/dx_1$) for the mixtures. Mixture 1 is stable, whereas mixtures 2 and 3 are unstable in that they have a tendency to form homogeneous clusters. From \cite{olsen2025new}.\label{fig7}}
\end{figure}

Olsen et al.\ \cite{olsen2025new} studied the intrinsic volumes of fluid mixtures using molecular dynamics.  It is well known that mixing a volume $V_1$ of water with a volume $V_2$ of ethanol produces a total volume $V<V_1+V_2$. If there is $N_1$ mol of species 1 and $N_2$ mol of species 2, the total number of mol are $N=N_1+N_2$.  The mol fractions are defined as
$x_1=N_1/N$ and $x_2=N_2/N$, so that
\begin{equation}
\label{eq10-1}
x_1+x_2=1\;,
\end{equation}
and the partial molar volumes are 
\begin{eqnarray}
\hat{\nu}_1&=&\left(\frac{\partial V}{\partial N_1}\right)_{T,p,N_2}=\nu+x_2\left(\frac{\partial \nu}{\partial x_1}\right)_{T,p}\;,\label{eq10-2}\\
\hat{\nu}_2&=&\left(\frac{\partial V}{\partial N_2}\right)_{T,p,N_1}=\nu-x_1\left(\frac{\partial \nu}{\partial x_1}\right)_{T,p}\;,\label{eq10-3}
\end{eqnarray}
where $\nu=V/N$, and we have that
\begin{equation}
\label{eq10-4}
\nu=x_1\hat \nu_1+x_2\hat \nu_2\;.
\end{equation}

We may also define {\it intrinsic molar volumes\/}
\begin{eqnarray}
\nu_1&=&\frac{V_1}{N}\;,\label{eq10-5}\\
\nu_2&=&\frac{V_2}{N}\;.\label{eq10-6}
\end{eqnarray}
Here $V_1$ and $V_2$ are the volumes that each species {\it actually\/} occupies.  It may e.g., be found by constructing a Voronoi tesselation of the volume $V$.  Each molecule is then assigned a volume, and $V_i$ is then the sum of the volume around each molecule of type $i$.  We have that 
\begin{equation}
\label{eq10-7}
\nu=x_1\nu_1+x_2\nu_2\;.
\end{equation}

Comparing Equations (\ref{eq10-4}) and (\ref{eq10-7}), we may now define a {\it co-molar volume\/} $\nu_m$ by the relations
\begin{eqnarray}
\nu_1&=&\hat \nu_1-x_2\nu_m\;,\label{eq10-8}\\
\nu_2&=&\hat \nu_2+x_1\nu_m\;.\label{eq10-9}
\end{eqnarray}

We show in Figure \ref{fig7} the co-molar volume measured in molecular dynamics simulations for three mixtures interacting in different ways. We introduce the notation\footnote{We use $w$ rather than $\mu$ which we used for the corresponding quantity in the porous media problem. This in order to avoid mixing it up with the chemical potential normally named $\mu$ in thermodynamics.}
\begin{equation}
\label{eq10-10}
w=\left(\frac{\partial \nu}{\partial x_1}\right)_{T,p}\;.
\end{equation}
For the same reasons that $\mu$ was the natural variable for $v_m$ in the porous media problem, $w$ is the natural variable for $\nu_m$ together with $T$ and $p$. We may write
\begin{equation}
\label{eq10-11}
\nu_m(T,p,w)=a(T,p)+b(T,p)w+\sum_{k=2}^\infty b_k(T,p)w^k\;.
\end{equation}
Analyzing the data in Figure \ref{fig7}, Olsen et al.\ \cite{olsen2025new} find that the three mixtures are close to having the parameters $b_k$ close to zero --- but for mixtures 2 and 3 {\it not\/} equal to zero. We note that mixture 1 is stable whereas mixtures 2 and 3 are not. It is remarkable that the behavior of the co-moving velocity shown in Figure \ref{fig6} and the co-molar volume in Figure \ref{fig7} are so similar.  It is encouraging that a notion imported from the non-thermal thermodynamics of immiscible two-phase flow in porous media carries over to ordinary thermodynamics and not only the other way.    

\section{Discussion and conclusion}
\label{conclusion}

This short and incomplete review was written to rid the reader of the notion that porous media, and especially flow associated with such structures, is messy and best left to engineers. In fact, immiscible two-phase flow in porous media seems to be an excellent example of a non-thermal system that borrows the mathematical structure of thermodynamics.  There are other examples of non-thermal thermodynamics, however, such as the non-thermal thermodynamics of powders developed by Edwards and coworkers,  \cite{edwards1989theory}, see also \cite{baule2018edwards}.

As for developing the theory for steady-state immiscible two-phase flow in porous media further, there are many directions to follow. The glassy flow phase discussed in section \ref{darcy_scale_phases} is one direction to pursue.  This is the flow phase that is most relevant for practical applications, and its properties should be investigated. From a fundamental point of view, the system offers a clean problem where experimental investigations may closely pair with theoretical and computational investigations.  

Another direction that we see as important is to move away from steady state flow. In terms of the thermodynamics-like formalism developed, the next step is to invoke non-equilibrium thermodynamics \cite{kjelstrup2017non}.  Rather than dividing the pressure $p$ into a wetting and non-wetting pressure, $p_w$ and $p_n$ as in Equation (\ref{eq2}), the driving forces would then be gradients in the agiture (and thus of $|\nabla p|$) and the flow derivative $\mu$.  There would be no need to introduce a capillary pressure function $p_c$. This is ongoing work.

As a last remark, we note that this short review is a first presentation of this material to the statistical physics community. Previous work along these lines have been published in journals typically not read by this community. Our hope is that we have managed to illustrate here that the problem of immiscible two-phase flow in porous media is well approachable as a problem within the realm of statistical mechanics.   

\section*{Acknowledgements}

We thank Jos{\'e} Soares Andrade Jr., Saman Aryana, Dick Bedeaux, Carl Fredrik Berg, Steffen Berg, Humberto Carmona, Erika Eiser, Eirik G.\ Flekk{\o}y, Daan Frenkel, Bj{\o}rn Hafskjold, Signe Kjelstrup, Anders Lervik, Marcel Moura, H{\aa}kon Pedersen, Thomas Ramstad, Subhadeep Roy and Per Arne Slotte for numerous discussions. 

This work was partly supported by the Research Council of Norway through its Centers of Excellence funding scheme, project number 262644, and by the European Research Council ERC (Grant Agreement 101141323 AGIPORE). 

\end{document}